\documentclass[
    ,final            
  ]
  {aipproc}

\layoutstyle{6x9}

\begin{document}

\title{Test of lepton universality and search for lepton flavor violation in $\Upsilon$(1S,2S,3S) decays at CLEO}


\author{Istv\'an Dank\'o\footnote{Talk presented at the Particles and Nuclei International Conference (PANIC05), October 24-28, 2005, Santa Fe, New Mexico, USA.} \\
(for the CLEO Collaboration)}{
address={Rensselaer Polytechnic Institute, 110 8th St., Troy, NY 12180, USA}
}

\begin{abstract}
We present the analysis technique and preliminary results of two ongoing analyses at CLEO  which put lepton universality and lepton flavor conservation to the test in $\Upsilon$ decays.
\end{abstract}

\maketitle


\section{Introduction}

Universality in the couplings of the three lepton flavors to the gauge bosons ($\gamma$, $Z$, $W$) and conservation of lepton flavor are fundamental assumptions of the Standard Model. Any observed deviation from this assumption would indicate the presence of new physics.

Utilizing the large data samples collected recently in the vicinity of the $\Upsilon$(nS) ($n=1, 2, 3$) resonances, the CLEO Collaboration is in a position to perform  precision test of these assumptions. Two examples are discussed here: one which tests lepton universality in $\Upsilon \to \ell^+ \ell^-$ decays, and another one which searches for the lepton flavor violating decay $\Upsilon \to \mu^+ \tau^-$.

The data used in these analyses were collected in $e^+ e^-$ annihilation at the Cornell Electron Storage Ring (CESR) with the CLEO III detector during 2001-2002 and consist of about 20 million $\Upsilon$(1S), 10 million $\Upsilon$(2S), and 5 million $\Upsilon$(3S) decays. Additional data collected at the $\Upsilon$(4S) and off the peak of each resonance are used to study background and systematics.

\section{Test of lepton universality in $\Upsilon \to \ell^+ \ell^-$ decays}

Decays of the $\Upsilon$ resonances to lepton pairs proceeds, in first order, via the annihilation of the constituent $b\bar{b}$ quarks to virtual photon: $\Upsilon \to \gamma^{\star} \to \ell^+ \ell^-$ ($\ell = e, \mu, \tau$). In case of the three $\Upsilon$ resonances below the open beauty threshold the decay to each $e^+ e^-$, $\mu^+\mu^-$, and $\tau^+\tau^-$ final state represents approximately 2\% of the total decay width \cite{PDG04}. (In sharp contrast, the leptonic branching fraction of the $\Upsilon$(4S), which decays dominantly to $B{\bar B}$ pairs, is about three orders of magnitude smaller than those of the lower lying resonances.) By measuring the decay rate of the $\Upsilon$(1S, 2S, 3S) to the different lepton flavors we can test the validity of lepton universality and probe possible physics scenarios beyond the Standard Model such as the existence of a non-standard light Higgs boson with a mass close to the $\Upsilon$ resonances \cite{Sanchis05}.

The figure of merit which we use to test lepton universality is the ratio of the efficiency corrected tau pair yield and muon pair yield, $R = N(\Upsilon \to \tau^+\tau^-)/N(\Upsilon \to \mu^+ \mu^-)$. The tau pair yield is measured using one prong tau decays ($\tau \to \ell\nu_{\ell}\nu_{\tau}$ and $\tau \to h(n\pi^0)\nu_{\tau}$) which represent about 75\% of total tau decays, and applying event selection criteria similar to those used in an earlier measurement of their branching fractions at CLEO \cite{CLEO2_tau}. The muon pair yield is determined using a method which closely follows that used in a recent measurement of the muonic branching fractions of the $\Upsilon$(1S, 2S, 3S) resonances \cite{Bmumu}.

The dominant background due to non-resonant production of the $\mu^+\mu^-$ and $\tau^+ \tau^-$ final state is estimated from off-resonance data collected $20-30$ MeV below each resonance and subtracted from the on-resonance yield after scaling by the luminosity and the $1/s$ dependence of the non-resonant cross section. We observe a clear excess on all three resonances which we attribute to the resonance decays. In contrast, no excess is observed at the $\Upsilon$(4S) in accord with its minuscule branching fraction.

The $\Upsilon$(2S, 3S) yields are also corrected for background from cascade decays to a lower $\Upsilon$ resonance (e.g $\Upsilon(2S) \to \pi^0 \pi^0 \Upsilon(1S)$) which subsequently decays into lepton pairs.
As a consistency check we have verified that the efficiency corrected $\mu^+ \mu^-$ yield is consistent with our previous measurement \cite{Bmumu} and that the measured $\mu^+\mu^-$ and $\tau^+\tau^-$ cross sections in off-resonance data are consistent with the theoretical expectation including higher order radiative corrections.

The preliminary ratio of the tau pair to muon pair yield for each resonance is
\[
\begin{array}{c}
  R \left[ \Upsilon(1S) \right] = 1.06 \pm 0.02 \pm 0.00 \pm 0.03 \\
  R \left[ \Upsilon(2S) \right] = 1.00 \pm 0.03 \pm 0.12 \pm 0.03 \\
  R \left[ \Upsilon(3S) \right] = 1.05 \pm 0.07 \pm 0.05 \pm 0.03
\end{array}
\]
where the errors are respectively due to statistics, uncertainty in cascade background subtraction, and other systematics that do not cancel in the ratio.
This measurement represents the first evidence for $\Upsilon(3S) \to \tau^+\tau^-$ decay which has not been observed previously, and a significantly improved branching fraction for the $\Upsilon(2S) \to \tau^+\tau^-$ decay which has been measured with a 100\% uncertainty \cite{PDG04}.
Although the ratio at the $\Upsilon$(1S) indicates a slight violation of lepton universality, it is too early to draw a definitive conclusion at this stage of the analysis.

\section{Search for lepton flavor violating decay $\Upsilon \to \mu^+\tau^-$}

Lepton flavor violating (LFV) decays are strictly forbidden in the Standard Model, and severely suppressed even in the presence of non-zero neutrino mass and neutrino mixing. On the other hand, many extensions of the Standard Model, such as GUT inspired models, supersymmetric models, and technicolor models, allow new scalar and vector bosons (leptoquarks, sleptons, Z') which can induce LFV decays \cite{Huo}. In addition, LFV decays can be also triggered by low scale quantum gravity effects (or other new physics at the TeV scale) or by extra spacial dimensions \cite{Silagadze}. These scenarios can lead to an appreciable $\Upsilon \to \ell \ell'$ decay rate with branching fractions up to $10^{-5}$.

The BES Collaboration has recently set an experimental constraint on $J/\psi \to \ell^+ \tau^-$ ($\ell = e, \mu$) decays \cite{BES} but the $\Upsilon \to \mu^+ \tau^-$ decay has never been tested. The purpose of this analysis is to search for LFV decays $\Upsilon(1S, 2S, 3S) \to \mu^+ \tau^-$ in the exclusive channel $\tau^- \to e^- \bar{\nu}_e \nu_{\tau}$ (charged conjugate modes are implicitly included in this analysis).

Candidate events are required to have exactly two tracks with opposite charge, one identified as an electron and the other as a muon. The overall signal detection efficiency is about 9\% (including ${\cal B} (\tau \to e \nu \nu) \approx 17\%$) based on a full Monte Carlo simulation of the detector. There are three possible background sources we have to deal with: tau pairs, radiative muon pairs (when the shower from the photon accidently overlaps with the projection of either muon in the calorimeter), and muon pair events with one muon decaying to electron in flight inside the detector.

In order to extract the signal we perform an unbinned extended maximum likelihood fit to the data with four components (one signal plus three backgrounds). The probability distribution function (PDF) used in the likelihood function is factorized into four terms describing the distribution of the beam energy scaled energy of the muon ($E_{\mu}/E_{\rm beam}$) as well as the electron ($E_{e}/E_{\rm beam}$) candidate, and the specific energy loss ($dE/dx$) and $E/p$ (the energy deposited in the calorimeter divided by the measured momentum) of the electron.
Data collected on the $\Upsilon$(4S) resonance are used to derive most of the PDF shapes for the likelihood function since the LFV decays of the $\Upsilon$(4S) are expected to be suppressed with the same factor as the leptonic decays.

The performance of the fitter and possible bias is extensively studied using toy signal Monte Carlo events merged to $\Upsilon$(4S) off-resonance data representing the same statistics as the expected background level in the $\Upsilon$(1S, 2S, 3S) data samples. The number of signal events returned by the fitter is consistent with the embedded signal Monte Carlo level and no significant bias is found. These preliminary studies indicate a signal sensitivity that corresponds to a branching fraction of about $10^{-5}$. Additional improvements in the technique are under study before attempting to fit the $\Upsilon$(1S, 2S, 3S) data.


\begin{theacknowledgments}
We gratefully acknowledge the effort of the CESR staff in providing us excellent luminosity and running conditions. This work is supported by the National Science Foundation and the U.S. Department of Energy.
\end{theacknowledgments}


\end{document}